\documentclass[twocolumn,showpacs,prl,preprintnumbers,amsmath,amssymb]{revtex4}

\usepackage{graphicx}% Include figure files
\usepackage{dcolumn}% Align table columns on decimal point
\usepackage{bm}% bold math

\newcommand{\bra}[1]{\ensuremath{\left< #1 \right|}}
\newcommand{\ket}[1]{\ensuremath{\left| #1 \right>}}

\begin{document}

\title{Towards quantum frequency combs:\\ boosting the generation of highly nonclassical light states by cavity-enhanced parametric down-conversion at high repetition rates}

\author{Alessandro Zavatta$^{1,2}$, Valentina Parigi$^{2,3}$, and Marco Bellini$^{1,3}$}
\affiliation{$^{1}$Istituto Nazionale di Ottica Applicata (CNR), L.go E. Fermi, 6, I-50125, Florence,
Italy; \\ $^{2}$Department of Physics, University of Florence, I-50019 Sesto Fiorentino, Florence,
Italy; \\ $^{3}$LENS, Via Nello Carrara 1, 50019 Sesto Fiorentino, Florence, Italy}
\email{bellini@inoa.it} \homepage{http://www.inoa.it/home/QOG}

\date{\today}

\begin{abstract}
We demonstrate the generation of multi-photon quantum states of light by cavity-enhanced parametric
down-conversion in the high-repetition-rate pulsed regime. An external enhancement cavity resonant
with the spectral comb of modes of a mode-locked pump laser provides a coherent build-up of the
pump intensity and greatly enhances the parametric gain without sacrificing its high repetition
rate and comb structure. We probe the parametric gain enhancement by the conditional generation and
tomographic analysis of two-photon Fock states. Besides its potential impact to efficiently
generate highly-nonclassical or entangled multi-photon states in many existing experimental setups,
this scheme opens new and exciting perspectives towards the combination of quantum and comb
technologies for enhanced measurements and advanced quantum computation protocols.
\end{abstract}


\maketitle
\section{Introduction}
Nonclassical and entangled multi-photon states are gaining ever increasing attention as fundamental
resources to probe the quantum world, to implement quantum communication and computation protocols, and
to develop promising future quantum technologies. In particular, multi-photon highly-nonclassical states
are key resources for quantum information processing using continuous variables~\cite{ourjoumtsev06}.
Furthermore, universal quantum computation based on photonic qubits recourses to multi-photon
entanglement to implement quantum algorithms with linear optics~\cite{prevedel07}, and multi-photon
entangled states are also crucial for applications in quantum metrology and
cryptography~\cite{nagata07,durkin02}.

Currently, most of the methods used to generate nonclassical and entangled multi-photon states are
based on spontaneous parametric down-conversion (SPDC) processes producing two-mode squeezed states
of the form
\begin{equation}
\ket{\psi}= \sqrt{1-|{\lambda}|^2} \sum_{n=0}^{\infty}\lambda^n \ket{n}_i \ket{n}_s
\end{equation}
in non-collinear signal and idler modes. The parametric gain $\lambda$ depends on the crystal
nonlinearity and is proportional to the amplitude of the pump field. In the low-gain limit, the
probability of producing entangled photon pairs scales linearly with the pump intensity, while the
generation of n-photon states depends on the n-th power of it. Dramatic enhancements of the
multi-photon production rates can thus be achieved even with relatively modest increases in the
pump intensity. Great efforts have been recently made in order to increase the parametric gain
$\lambda$ and thus enhance the weight of multi-photon contributions to the emission, both in the
continuous-wave and in the pulsed regimes. While, in the former case, optical parametric resonators
are adopted to overcome the limits connected to the low intensity of cw pumps (see for example
Refs.~\cite{neergaard-nielsen06,takeno07,vahlbruch08}), ultra-short and ultra-intense pump laser
sources are also being actively used~\cite{kiesel07,lu07,demartini05}, albeit with complex
amplified laser systems and at low repetition rates.

Another recent revolution in modern physics has been triggered by the advent of femtosecond
frequency combs. Their invention has re-defined the entire field of precision measurements,
demonstrating an enormous potential in accurate frequency determination and space-time
positioning~\cite{udem02,li08}. Even if frequency-comb techniques have so far belonged to the realm
of classical physics, it is foreseen that the ever increasing accuracy in comb stabilization will
soon get close to measurement quantum limits and quantum-enhanced comb measurements will then
require the generation of nonclassical light with a comb structure~\cite{fabre08}. From an entirely
different perspective, the possibility of nonlinearly coupling the huge number of modes of a
frequency comb has also been recently proposed as a promising way to realize large and arbitrarily
scalable cluster states for one-way quantum computing~\cite{pfister08}.

It is therefore of high interest to transpose the many advantages offered by frequency combs to the
quantum domain. This clearly requires the generation of highly-nonclassical and entangled states
possessing a comb spectral structure, hence in high-repetition-rate pulsed schemes. Although a few
experiments have already succeeded in producing nonclassical states from pulsed laser systems at
high repetition rates~\cite{science04,pra04,science07}, only very low parametric gains are normally
obtained in such cases.

Here we demonstrate the use of an external enhancement cavity to coherently boost the intensity of
pump pulses and thus greatly increase the SPDC gain. Differently from previous approaches, this is
now obtained without additional costs in terms of laser power and, more importantly, without
compromising the high repetition rates required for comb applications. We verify the generation of
multi-photon nonclassical radiation by producing and tomographically analyzing two-photon Fock
states.

Although resonant enhancement cavities are quite common in the continuous-wave regime, there are
only a few applications in combination with pulsed laser sources. Early experiments with picosecond
pulses demonstrated high-efficiency generation of low-order harmonics~\cite{persaud90,watanabe94};
recent approaches have concentrated on the generation of high-order harmonics in a gas jet at high
repetition rates~\cite{gohle05,jones05} in an effort to transpose the femtosecond frequency-comb
structure of the pump laser to the extreme ultraviolet. Indeed, a pulsed enhancement cavity is
essentially made of a ring resonator whose longitudinal mode structure exactly matches the comb of
modes (spaced by the pulse repetition rate) of the mode-locked source laser. This is accomplished
by carefully adjusting and locking the external cavity length to that of the laser cavity. In the
time domain this condition is seen to give rise to a constructive interference between the pulse
circulating in the enhancement cavity and those coming from the laser. The coherent addition of the
energy from many successive pulses of the laser pulse train can thus result in a significant
build-up of the intracavity energy.

\section{Experimental setup}
\begin{figure}
\includegraphics[width=85mm]{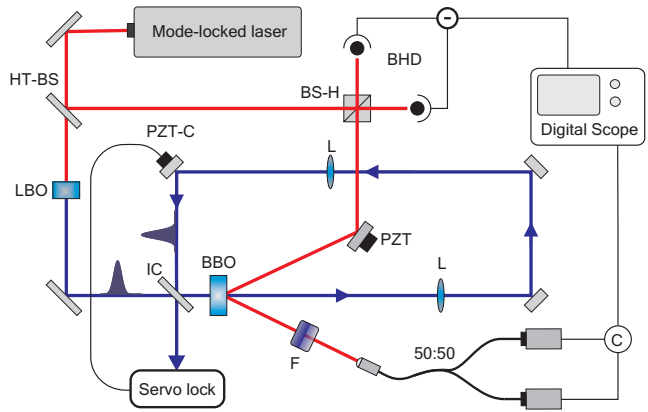}
\caption{\label{fig:setup} (color online) Experimental setup. The pump enhancement cavity (a simplified
version is shown here for clarity) is built around the SPDC crystal (BBO). Along the idler channel two
on/off detectors are connected to a 3-dB single-mode fiber coupler after a series of spectral and
spatial filters (F). The high-transmission beam-splitter (HT-BS) is used to extract a small portion of
the main laser source to be used as a reference beam (local oscillator) in the the balanced homodyne
detector (BHD) after mode matching optics (not shown). The phase between the local oscillator and the
analyzed state is adjusted by a mirror mounted on a piezoelectric stage (PZT). Once one of the two (or
both) on/off detectors clicks, the homodyne signal is acquired by a digital scope and then analyzed.}
\end{figure}

The experimental setup shown in Fig.~\ref{fig:setup}
consists in a second harmonic generation crystal (LBO) which generates radiation at 393 nm from a
1.5-ps mode-locked laser with a repetition rate of $\mathcal{R}=82$ MHz. The UV beam pumps a 3 mm
type-I BBO crystal and produces SPDC into well-defined idler and signal spatial modes. In order to
conditionally generate photon Fock states, a pair of on/off detectors (single photon counting
modules SPCM AQR 14) is placed after a $50\%$ beam splitter in the idler channel. When one detector
or both click, a single- or two-photon Fock state, respectively, is prepared in a well-defined
spatiotemporal mode along the signal channel. In principle, using this setup it is possible to
prepare Fock states of any order $n$ depending on the preparation measurement performed on the
idler channel. Up to now, Fock states with $n=1, 2$ only~\cite{lvovsky01,pra04,ourjoumtsev06:prl}
have been generated and characterized by quantum tomography, due to the very low gain in the SPDC
process. In this configuration the expected rates of single and two-photon state production
($\mathcal{R}_1$ and $\mathcal{R}_2$, respectively) are simply related as
$\mathcal{R}_2=\mathcal{R}_1^2/2\mathcal{R}$, implying that any enhancement in the single-photon
production rate scales quadratically in the two-photon one.

After the preparation of a given state, balanced homodyne detection is performed in the signal channel
by mixing the signal state with a strong reference beam called local oscillator using a $50\%$ beam
splitter (BS-H). The outputs of the beam splitter are then detected by proportional photodiodes
connected to a home-made wide-bandwidth amplifier~\cite{josab02}. From the acquisition of many homodyne
data for states prepared in the same way one can obtain the quadrature distribution of the state for a
given phase between local oscillator and the signal. A complete set of quadrature distributions at
different phases allows one to reconstruct the density matrix and Wigner function of the analyzed signal
state.

An enhancement cavity with a length of 3.6 m, corresponding to the 12 ns time delay between
successive pulses from the laser, has been built around the SPDC crystal. It uses 7 low-loss plane
mirrors with reflectivity $R\approx 99.95$\% for the pump pulses at 393 nm. Two lenses (L) with a
focal length of 600 mm are carefully positioned in the cavity in order to produce a beam waist of
about 250 $\mu$m inside the SPDC crystal. In this configuration the measured cavity losses
(indicated by $1-R_m$, where $R_m$ is the overall effective cavity reflectivity) amount to about
7\%, mainly due to the residual reflections and absorptions on the crystal and the lenses. An input
coupler (IC) with a reflectivity $R_{i}=90$\% is used (under-coupled cavity configuration). A
portion of the beam reflected from the input coupler is used to lock the cavity to the resonance
peak by using the method proposed by H\"{a}nsch and Couillaud~\cite{hansch80}. The expected cavity
power enhancement,
\begin{equation}
\mathcal{E}= \frac{1-R_i}{(1-\sqrt{R_i R_m})^2},
\end{equation}
is then calculated to be $\mathcal{E}=14$, with a cavity finesse of $\mathcal{F}=35$, which is in very
good agreement with the measured one.

Differently from the schemes used for intracavity second harmonic
generation~\cite{persaud90,watanabe94}, here pump depletion plays no role in limiting the
enhancement factor by losses, because of the very low parametric gain that allows an almost
complete recycling of the pulse energy after interaction with the crystal. Moreover, the use of a
non-collinear SPDC configuration does not impose an output coupler for the down-converted light,
and thus eliminates another important source of losses (as those experienced when trying to couple
XUV light out of the cavity in \cite{gohle05,jones05}). Finally, the use of a picosecond pulse
source allows us to avoid the problems connected to intracavity dispersion when working with
ultrashort femtosecond pulses. However, these will have to be taken into account when dealing with
femtosecond frequency combs, since carrier phase stabilization is only possible in the ultrashort
pulse regime.

In order to verify the production of nonclassical multi-photon states we first check the generation
efficiency of single-photon Fock states by using a single on/off detector in the idler channel. The
preparation rate in this case is 5.8 kHz, to be compared to the 500 Hz of previous experiments
performed by our group with the same pump power~\cite{science04,zavatta06,science07}. The
single-photon production rate is thus enhanced of a factor of about 12, in good agreement with the
expected increase of pump energy. Another peculiar advantage of using an enhancement cavity is that
it works as a Gaussian spatial filter for the pump pulses. This has allowed us to simplify a part
of the setup and further decrease the overall system losses by removing the pinhole-based spatial
filtering of the second harmonic pump light that was present in our previous
experiments~\cite{note}.

\section{Generation of two-photon Fock states}
\begin{figure}
\includegraphics*[width=85mm]{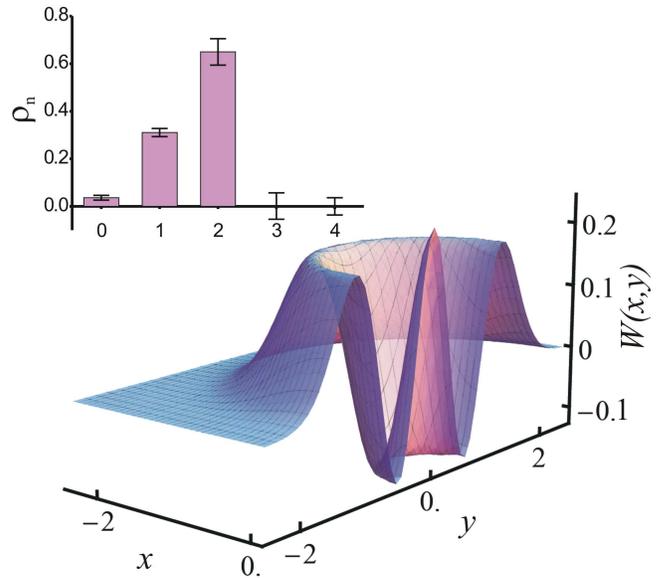}% Here is how to import EPS art
\caption{\label{fig:fock2} (color online) Experimentally reconstructed density matrix diagonal
elements and corresponding Wigner function for the two-photon Fock state generated by
cavity-enhanced SPDC.}
\end{figure}

Using the enhancement cavity we are now also able to generate two-photon Fock states at a
sufficient rate without any increase in the pump laser power. We have acquired about 7000
quadrature measurements with a mean rate of $0.14\pm0.05$ Hz in about 14 hours of experimental run.
This rate represents a 150-fold increase over the exceedingly low value found for a single-pass
configuration and, as expected, the two-photon enhancement factor is approximately the square of
the one measured for the single-photon case. In Fig.~\ref{fig:fock2}
we show the reconstructed density matrix elements and the resulting Wigner function for the
two-photon Fock state. The maximum likelihood method~\cite{lvovsky04,hradil06} has been used to
retrieve the 5 diagonal elements of the density matrix and the contribution of the inefficient
detection ($\eta_d=0.67$) has been taken into account in the reconstruction procedure. A clear
central peak and a negative ring region around the origin of the quadrature axis space are evident
(and are still present even without correcting for detection losses) in the Wigner function, the
latter being a sign of the highly nonclassical character of the generated state. From the
reconstructed elements of the density matrix (inset of Fig.~\ref{fig:fock2}) the contribution of
the residual $\ket{0}\bra{0}$ and $\ket{1}\bra{1}$ terms is still evident. These are due to
residual impurities in the state preparation, which contribute as losses with $\eta_{p_2}=0.81$. On
the contrary, no higher-order contributions are visible in the density matrix.

\section{Discussion and conclusions}
\begin{figure}
\includegraphics[width=85mm]{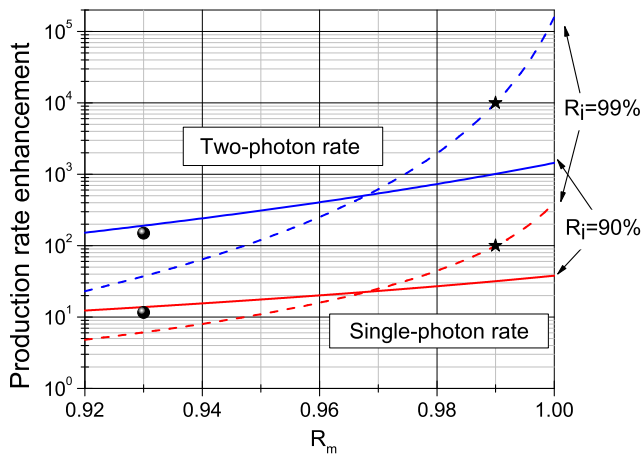}% Here is how to import EPS art
\caption{\label{fig:cavity} (color online) Plot of the cavity-enhanced production rates for the
single- (red curves) and two-photon (blue curves) Fock states. The horizontal axis represents the
overall effective reflectivity $R_m$ of the cavity, and two curves are calculated for different
reflectivities $R_i$ of the input coupler ($R_i=99\%$ dashed curves; $R_i=90\%$ solid curves).
Current experimentally-measured enhancement factors are indicated by solid circles corresponding to
$R_m=0.93$ and $R_i=90\%$. The expected enhancements for an optimized, impedance-matched, cavity
with $R_i=R_m=0.99$ are indicated by stars.}
\end{figure}

It is worth noting that cavity losses can be easily reduced to below 1\% by improving the design of
the enhancement cavity: this involves the use of fewer mirrors and of concave ones to replace the
lenses, and the application of an ultra-high-quality anti-reflection coating to the SPDC crystal
faces. In such a case, and in the impedance matching condition (with $R_i=R_m=0.99$), we can expect
a cavity finesse of $\mathcal{F}=300$ and an enhancement factor up to about $\mathcal{E}=100$. The
resulting increase in the parametric gain would thus reflect in a $10^4$ enhancement with respect
to the two-photon preparation rate without the cavity (see Fig.~\ref{fig:cavity}).
The enhancement-cavity approach would thus allow us to achieve the same rates as obtained by
Ourjoumtsev et al.~\cite{ourjoumtsev06:prl} with a cavity-dumped femtosecond laser working at a
repetition rate which is a factor of $10^2$ smaller than ours.

In conclusion, we have demonstrated a simple technique to greatly enhance the gain of pulsed parametric
down-conversion and the production of nonclassical multi-photon states while preserving the high pump
repetition rate and its comb spectral structure. A picosecond pump enhancement cavity built around the
SPDC crystal is shown to provide an enhancement factor of about 15 (see~\cite{note}) in the production
rate of single-photon Fock states, and an increase of more than two orders of magnitude in the rate of
two-photon state generation. With a further slight reduction of cavity losses the enhancement factor
would still greatly increase. Furthermore, if issues connected to cavity dispersion are properly taken
into account, the use of intracavity SPDC with phase-stabilized femtosecond pump pulses will provide
much higher gains and a stable frequency comb structure.

The extension of this approach with the adoption of a synchronously-pumped optical parametric oscillator
configuration, where collinear degenerate parametric down-conversion is used in a cavity resonant also
for the signal and idler modes, would give access to multimode squeezing~\cite{fabre06} and multipartite
continuous-variable entanglement as recently proposed for the realization of cluster states for one-way
quantum computation~\cite{pfister08,pfister07}.

We believe that this technique will have a strong impact in a more widespread production of
highly-nonclassical states, giving the opportunity to investigate higher-dimensional Hilbert spaces and
multi-photon entanglement even with modest available pump powers. Moreover, it will help combining two
of the most intriguing and promising avenues in modern physics, opening the way towards quantum-enhanced
frequency-comb technologies and to appealing schemes for quantum information processing.

This work was partially supported by Ente Cassa di Risparmio di Firenze and by CNR-RSTL Projects.

%\bibliography{Fock_bib}

\begin{thebibliography}{30}
\bibitem{note}{By also considering the diminished pump losses obtained by eliminating the pinhole-based,
mode-cleaning spatial filter, an overall enhancement of the single-photon production rate of about 15
(corresponding to a fourfold increase of the SPDC gain) with respect to the single-pass configuration
has been reached.}
\end{thebibliography}

\end{document}